# Computational Rational Engineering and Development: Synergies and Opportunities

Ramses Sala[1]

[1] Technische Universität Kaiserslautern,
Department of Mechanical and Process Engineering, 67663 Kaiserslautern, Germany
`sala@rhrk.uni-kl.de`

**Abstract.** Research and development in computer technology and computational methods have resulted in a wide variety of valuable tools for Computer-Aided Engineering (CAE) and Industrial Engineering. However, despite the exponential increase in computational capabilities and Artificial Intelligence (AI) methods, many of the visionary perspectives on cybernetic automation of design, engineering, and development have not been successfully pursued or realized yet. While contemporary research trends and movements such as Industry 4.0 primarily target progress by connected automation in manufacturing and production, the objective of this paper is to survey progress and formulate perspectives targeted on the automation and autonomization of engineering development processes. Based on an interdisciplinary mini-review, this work identifies open challenges, synergies, and research opportunities towards the realization of resource-efficient cooperative engineering and development systems. In order to go beyond conventional human-centered, tool-based CAE approaches and realize Computational Intelligence Driven Engineering and Development processes, it is suggested to extend the framework of Computational Rationality to challenges in design, engineering and development.

## 1  Introduction and motivation

Advances in computer technology and computational science have provided crucial tools to aid the engineering and realization of a wide variety of mechanical structures and systems [1–3]. Examples of influential tools are: the geometrical modeling by means of Computer-Aided Design (CAD) [4, 5], the simulation and analysis of virtual prototypes using Computer-Aided Engineering (CAE) tools [6], and automated machining using Computer-Aided Manufacturing (CAM) [1, 7]. The increase in computational engineering capabilities, however also led to a progressive increase in the complexity of processes and products, which poses a massive challenge for modern industrial engineering [8]. After the 1990s, a paradigm shift in engineering design was expected due to the developments in the fields of Computational Intelligence (CI), Soft Computing (SC), Machine Learning (ML), and AI [9]. But, despite that the capabilities of computational tools for specific tasks in the engineering process have improved exponentially, the structure, organization, and paradigms of the overall design, engineering and development processes have been adapted only modestly [10]. Many of the visions and expectations on automated engineering and development systems formulated in the early literature [11, 12] have not yet been realized [13]. To address the challenges of increasing complexity in product design, engineering and development [8], new paradigms and research frameworks might be needed [10].

*How to effectively realize AI technologies and intelligent systems that can enable improvement and automatization of industrial design, engineering, and development processes?* To analyze and discuss the many aspects of this quest, seminal classical works as well as recent results from the fields of Systems Engineering, Computer Science, Computational Mechanics, Uncertainty Quantification, and Operations Research, Cognitive and neuroscience, are reviewed with a focus on intersections related to the understanding and automation of problem-solving and decision making in design, engineering, and development. Based on the presented interdisciplinary mini-review progress, open challenges, synergies, and perspectives on directions for further research are identified, formulated, and discussed in the following sections.

## 2 Recent and past perspectives on computer systems for automation of engineering and development

Relatively soon after computing machines or early computers became available for non-military purposes, they were applied in the development process of various engineering products [14]. Besides the obvious applications of computers for calculations, more revolutionary ideas, concepts, and theories for computer-aided design systems were established in the 1960s [4]. In particular, the development of graphical human-machine communication interfaces enabled new possibilities for Computer-Aided Design (CAD). Around the same time, also new computational methods to model, simulate and optimize the response of complex systems and structures were developed [15, 16]. Other early developments relevant to engineering automation were general problem-solving programs [17] and expert systems [18]. These and other seminal works initiated the research and development, which eventually resulted in the wide variety of Computer-Aided technologies (CAX) [19] that provide today's state-of-the-art tools for engineering development [5, 20–24].

In 1960, early visionary perspectives related to automation of the engineering process were presented in [11]. The paper described expected developments towards intuitive man-machine cooperation and interaction technologies that would enable computers to facilitate the problem formulation and decision-making processes for complex engineering endeavors. The described targets aimed to go beyond mechanical extensions and mere automation of prescribed tasks, resulting in a man-computer symbiosis, that would enable thinking capacities as no human brain has ever thought. *"One of the main aims of man-computer symbiosis is to bring the computing machine effectively in to the formulative parts of technical problems"* [11].

Other visionary concepts of Intelligent Computer-Aided Engineering (ICAE) were described in [12]. The conceptual ideas were presented as a roadmap towards long-term targets for the development of computer programs or partners that could capture engineering knowledge to assist engineers in the engineering design, realization maintenance, and operation of engineering products. Some of the identified concepts required to achieve ICAE were: broad domain models, layered domain models, routine design, functional descriptions, qualitative simulation, and communication [12]. Furthermore, the importance of developing methods for the hierarchical decomposition of physical problems and qualitative physics models to approximate the responses of the systems and subsystems were highlighted. Also, the necessity for long-term research commitments to go beyond incremental progress was emphasized.

In a perspective paper [25] identifying general open challenges in the field of AI and CI, also important aspects and challenges of importance to the automation of engineering and design were identified. Human intelligence, and the type of intelligence measured by the Turing test, is very multidimensional. These dimensions of intelligence are often considered separately, and many systems can only be considered "partially intelligent". An important observation was that no adequate test or performance measures to quantify utility and integration in AI systems of partially intelligent agents are available [25]. Furthermore, it was highlighted that: *"For an artifact, a computational intelligence, to be able to behave with high levels of performance on complex intellectual tasks, perhaps surpassing human level, it must have extensive knowledge of the domain."* [25].

The perspective on intelligent machines in the context of engineering design [10] identified that most modern computer-aided design tools are still essentially extensions of engineering and practices going back more than two centuries. It was also was highlighted that: *"Today's innovations in robotics, advanced materials and additive manufacturing require newer and more creative design processes, enabling an entirely new kind of Arsenale — an Arsenale in which computers work as our creative partners"* [10]. From that perspective, computer-augmented design was identified as a next step beyond merely computer-aided design.

The article in [26] on cognitive AI systems provided a discussion on important bottlenecks and topics for further research targeting human-level functionality AI [26]. Many AI or CI systems that intend to aid humans cognitively can be categorized as: (i) Cognitive prosthesis or (ii) Cognitive orthotics. The aim of cognitive prosthetic systems is to operate independently before human supervision. An example of a prosthetic system is for example, machine translation such as Google translate. Although it generally needs human modifications, it is considered a cognitive prosthetic because it operates fully independently before human interaction is needed. Cognitive orthotic systems are characterized by the intent to enhance human capabilities and require human-machine interactions. An important build-in quality ceiling of such systems is the communication with humans [26]. The work pointed out that *"In order to burst through the quality ceiling and move toward comprehensive applications that are more like intelligent agents than mechanistic automata, the field must readdress newly available theories and methods, the development of systems featuring human-inspired computational models."(p7.* [26]*).*

Relatively recent strategic research initiatives and trends such as "Industry 4.0 "[27] and "Made in China 2025" have a strong emphasis on manufacturing and focus less on the engineering design and product development processes [28]. Although these new perspectives and projections on the future of industrial automation lean towards cyber-physical-systems and advanced human-machine interactions, those visionary concepts however still paint a rather human-centered picture in the execution (see also [29]).

Why have intelligent systems as envisioned in [11, 12] with capabilities beyond the current CAX tools not been realized yet [10, 30], despite all progress and advances in computation, simulation, ML, CI, and AI? Based on the articles discussed in this section, several trends, open issues, and obstacles towards automation in engineering and development can be identified and summarized:

1. There seems to have been a trend to focus on human-centered engineering development paradigms and automation approaches such as tool-based systems, cognitive orthotics, and man-computer symbiosis [26, 29].
2. Human-machine communication is still a bottleneck in current intelligent systems for automation in engineering and design [26, 29, 31].
3. Improved domain descriptions and models of the various agent tasks, environments, and resources in engineering and development processes are required [25, 26, 31].
4. The progress and success of AI for narrow tasks seems to have diverted the attention from long-term high-level goals on the automation of complex design and engineering processes towards the many lower hanging fruits in the field of AI and automation [10, 12]. To break the quality ceiling, research that targets intelligent systems with higher-level capabilities is necessary [26].

In the following sections, general and domain-specific aspects central to the research and development of intelligent systems for design, engineering, and development are reviewed and discussed in order to highlight promising directions and areas for future research on automation and decision-making in engineering development.

## 3 Computational Rationality in Engineering Development

### 3.1 Domain characteristics: problem-solving and decision-making in the context of industrial design, engineering, and development

Industrial engineering and development are often associated with the resulting technological products and impact on our environment. The resulting technological products are, however, only the tip of the iceberg of the engineering development process. Industrial engineering not only involves the design and engineering of a technological product, but it also involves the planning and development of the processes and facilities involved in material extraction, manufacturing, control, maintenance, and recycling during the product life-cycle stages. Furthermore, not only the final product and the involved production processes, but also the product development process itself (the organization and structuring of all the involved activities), needs to be established and realized in a way that satisfies requirements on performance, quality, cost, sustainability, and other operational aspects. The following sections review aspects related to intelligence and rationality in the operational and executive processes of engineering development. For interesting aspects beyond rationality related to sustainability, and ethics of the product and process objectives and requirements defined and set by humans is referred to [32–34].

**Process and Problem Complexity.** The development of engineering products and systems can involve thousands of people over several years. Increasing complexity is one of the biggest challenges in engineering design and modern product development [8]. Many products are becoming increasingly complex due to the integration and blending of various state-of-the-art technologies, such as composite materials, smart materials [35], and distributed control systems. Large-scale concurrent engineering on complex projects involves many tasks, sub-problems, various types of uncertainties [36, 37], decision-making based on incomplete information, and a dense web of information flows and interdependencies [38]. The engineering and product realization process of complex products has itself become a complex system, one that could be described as "organized complexity" [39].

**Hierarchical Bounded Rationality.** From an industrial engineering perspective, the development of a product generally involves a composition of many interdependent decisions and tasks in a complex hierarchical structure, which all need to be solved using a common resource budget that needs to be allocated over all activities to achieve a common objective. The core challenge in industrial engineering is to organize and address the many sub-tasks in order to realize the overall objectives using only limited information, knowledge, and other resources. The industrial engineering context thus poses a scenario of Bounded Rationality (BR) [40] at the level of individual tasks as well as at the level of the organization [41]. Although the environment and policies of agents dealing with technical decision-problems and organizational problems might be very different, the general concepts from the framework of Computational Rationality (CR) [42] could be used to target further progress in understanding and automation of engineering activities. Although there has been research on hierarchical decision-making [43–45], synergies with concepts from BR and CR for hierarchical decision problems in engineering development seem rather underexplored.

**Uncertainties in expected utility and resource use.** Challenging aspects in design, engineering, and development processes are the errors and uncertainties involved in the estimation of the system response behavior before its realization [36]. Although by means of virtual-prototyping and simulations, the response of physical systems can be approximated, reliable estimations for the simulation accuracy and effort are still difficult to obtain, especially for nonlinear systems. While there has been substantial progress in the areas of error estimation [46], uncertainty quantification [47–49], Global Sensitivity Analysis [50], and related areas [51] in academic settings, the application of these methods in industrial settings are still relatively rare. Therefore, further work targeting deeper integration of uncertainty quantification in industrial engineering and development processes would be beneficial.

**Non-rational design criteria and problem specification.** Although the postulation of the design objectives, requirements, and targets are often considered as non-rational [52], many evaluation criteria used in engineering are heuristics in disguise. The high-level, truly non-rational designer preferences often require a translation or reformulation of lower-level technical goals, requirements, and objectives. The activities related to formulating and specifying technical objectives and requirements at various levels of detail are related to the value alignment problem [53] and reward specification in reinforcement learning (RL). Hierarchical (heuristic) sub-problem approximations and approximate rewards or utilities could play a role in problem-solving [54], [55]. Further development of approaches that combine data-mining and simulation workflows (e.g. [56, 57]) could also improve the formulation and specification of partial approximate design evaluation criteria, utility, and reward functions. Besides data and information mining to extract useful design specifications, also effective languages are required. Although several modeling methods and languages have been presented, they are still deficiencies in generality for requirement specification [58]. The work in [59] indicated that it is even not clear how to evaluate and compare the different modeling methods and languages. Relatively recently also reward modeling techniques for RL have been developed, which can efficiently learn from (interactively communicated) human preferences for those decision problems where the evaluation criteria are difficult to specify in formal languages [60]. Since in engineering and development, not only the physical implementation of the systems but also the specification of goals, requirements, and targets can be complex, further work in these directions is required.

### 3.2    Interdisciplinary opportunities and synergies

**Computational Rationality**. "*A rational agent is one that acts so as to achieve the best outcome or, when there is uncertainty, the best expected outcome.*"[53]. Rational agents thus seem the ideal candidates for many activities, including decision making and problem-solving in engineering and development. Because in an industrial engineering development setting, knowledge time and other resources are limited, while there are many tasks and decision problems, agents must decide and act under conditions of Computational Rationality (CR). In a nutshell: the challenge is not only what to decide, but also how to decide, given the available resources. The meta-level decisions about resource allocation and method or policy selection in agent-based bounded rational decision making can be based on metareasoning using metalevel models or on heuristic decision policies [42]. The framework of CR [42, 61] aims to unify the fields of AI, cognitive science, and neuroscience in order to exploit synergies between the fields. The goal of CR is: *"Identifying decisions with the highest expected utility, while taking into consideration the cost of computation in complex real-world problems in which most relevant calculations can only be approximated"* [42]. This is also relevant in the context of understanding, formalizing, improving, and eventually automating engineering development processes. The perspective of understanding intelligence as computational rationality is in principle domain agnostic and open to consider human, natural, as well as artificial systems and activities.

**Neuroscience and Cognition**. Engineering and development involve decision-making and problem solving under limited knowledge, time, and other resources. In the framework of CR [42], two directions to address such problems are model-based metareasoning and the application of heuristic methods. Limited resources can make detailed metareasoning or formal methods unfeasible and can justify the use of heuristics for artificial as well as human agents [42, 62]. In [63, 64] systematic errors and biases in common human heuristics and interesting insights on fast heuristics and slow reasoning were identified. The work in [65, 66] highlighted the importance in human agents of matching patterns in the environment with decision heuristics. In [67] various Bayesian-based approaches to build intelligent systems using reverse-engineering of human cognitive functionalities and development were reviewed. This work emphasized the importance of language and hierarchical flexible structured data representations for cognitive capabilities such as abstraction and generalization [67]. In [68], concepts of BR are used and combined with set-based design, meta-modelling and multi-objective optimization to improve decentralized design problems. Investigations in [69] on a human grandmaster chess player indicated the importance of recognition compared to look-ahead search based on investigations on human experts. The theory of Ecological Rationality formalizes that the rationality of a decision policy depends on the circumstances [70]. This conclusion matches in spirit with the results of the No Free Lunch (NFL) theorems [71, 72]. Improved understanding of decisions and meta-decisions in human cognitive processes and other aspects of psychology could contribute to insight and development of computational methods in AI, engineering and science [62, 73–75], and maybe also vice versa.

**Design and Engineering Science.** Design and Engineering can benefit from strategic, systematic, and scientific approaches [76, 77]. In order to use computers and computational methods to solve design and engineering problems, it could help to establish formal (mathematical) descriptions of the problems or tasks of interest [78, 79]. Aspects related to creative design and problem-solving in the development process can be transformed in constraint satisfaction, optimization and search problems using Formal Design Theory (FDT) [80]. The use and extension of FDT and other formal design approaches (see also [81]) could support the frontiers of research on the automation of engineering design. Surveys on various theories and process models of engineering design have concluded: that presently no single model can address all issues and that different models may be useful for different situations [38, 82]. There are still many aspects of design and engineering which have not yet been rigorously formalized and which thus still pose open challenges and opportunities. Education and further research on general formal design theories and engineering science seem therefore of crucial importance for automation of engineering design and development.

**Computational physics and uncertainty quantification.** To make predictions and inferences on systems and processes, numerical models and simulations can be used. Computational Physics and Mechanics based models are commonly used in robotics, control and computational engineering of physical systems. In [83], a differentiable physics simulation was presented, which enabled the use of gradient-based methods in the control and optimization of physical/mechanical systems. A new approach to use physics simulations combined with multi-level path planning in the context of robotics was described in [84]. Conversely, methods to learn and infer physical principles from data have been presented in [85, 86]. The accuracy of physical models and simulations in general is limited due to errors and uncertainties and requires tradeoffs w.r.t accuracy and computational effort. Important approaches to address and investigate these accuracy limitations are: Validation and Verification (VV) [87], and Uncertainty Quantification (UQ) [48, 88], and Global Sensitivity Analysis [50] approaches.

**Optimization and Control.** Many sub-tasks and design problems in engineering can be formulated as optimization and control problems. In combination with physics engines or numerical models and simulations, the approximate representation of the properties and behavior of physical systems or processes can be optimized with respect to specified design objectives and constraints. The simulations and responses involved are, however, often relatively complex and computationally non-trivial, such that the selection and tuning of effective optimization algorithms is difficult. Optimization and automated design approaches and workflows have been developed and investigated for applications as: topology optimization and generative design of structures [89, 90], circuit design [91, 92], Elevator Systems [93] bioelectrochemical systems [94], automotive control actuators [95] and electric vehicle transmissions [96]. These examples demonstrate the use and potential of automated Modeling Simulation and Optimization (MSO) workflows for specific applications of industrial relevance. General frameworks for MSO-workflows that include automated agents for decisions regarding modeling accuracy, model parameterization, algorithm selection, and computational resources, are however still lacking, and seem a promising direction for further research. In the context of massive complex software systems, the use of Bayesian Optimization was proposed relatively recently in [97]. In [97] Bayesian Optimization was recognized as a powerful tool to address the many distributed design choices, and a key ingredient to take humans out of the loop in the development of complex software systems. The Bayesian perspective also highlights the importance of model selection, the consideration of uncertainty, and learning or model updating.

When design problems are formulated as true Black-Box optimization or search problems over finite search spaces, the NFL theorems [71, 72] apply. These theorems imply that no universally superior algorithms exist when performance is averaged over all possible problems. Thus, the remaining quest is to match specific problem classes of task-environment-resource combinations with specific efficient policies or algorithms. This, in turn, highlights the importance of: a) problem characterization and categorization (or fitness landscape analysis) [98–101]; b) systematic and generalizable optimization algorithm benchmarking [102–104]; c) algorithm performance analysis and selection [99, 105–107]. While there has been increasing interest towards algorithm selection for black-box optimization problems in a general context [99, 108] as well as for simulation-based engineering applications [100, 109], there are still many open challenges of scientific and practical relevance related to optimization algorithm benchmarking, selection and analysis [104, 106]. The extended process-perspective of optimization to the meta-level also highlights the need for optimization algorithm performance measures that go beyond fixed-budged and fixed-target performance evaluation criteria, to also include measures that can be used in dynamic hierarchical settings. Such settings involve decisions regarding method selection and resource allocation, which require more complex performance measures involving estimations of the expected utility per resource use, also considering the uncertainties.

**Operations Research and Systems Engineering.** Although not always directly targeted at computer-based automation, interesting methods and strategies to manage the design of complex systems have been developed in the fields of Engineering Management, Operations Research, and Systems Engineering, which could also benefit the automation of engineering and development processes [110–112]. One research direction towards a general approach to manage complexity in systems engineering is Model-Based Systems Engineering [113], there are, however, still many open challenges, and further work is needed to close the gaps between theory and implementation [114]. One of these challenges in to establish models that do not only estimate the expected results but also quantify the uncertainties. One interesting contribution in this research direction is the concept of Experimentable Digital Twins (EDT) [115]. The idea is to establish communication between virtual twin models, which represent the data, functions, and capabilities of real objects or processes, in networks of communicating EDTs on a system level, in order to realize complex control systems. In [116] the potential applicability of RL and ML in the domain of Systems Engineering was discussed, and it was concluded that further work in this promising direction was recommended.

**AI and CI** cover many areas of high relevance to intelligent systems in general [53, 117]. The following sub-sections highlight recent progress from various sub-fields of specific importance for automation in design and engineering processes.

*Automated Software development.* Interesting automated software testing and design approaches have been presented that could contribute to the automation of engineering and development of physical products [118–121].

*Agent and Multi-Agent Models, Systems and Control.* Complex processes can be modeled and controlled by means of agent-based and multi-agent models and systems. [122–125]. Multi-agent based models and systems can be combined with systematic management and systems engineering approaches [123, 126].

*Knowledge-based systems* for applications in Engineering, often referred to as Knowledge-Based Engineering (KBE), is another approach to capture, store and reuse information that could be used in engineering and development [127, 128]. A review of developments and open challenges for KBE systems is presented in [23].

*Robotics and control.* In the research field of evolutionary robotics, several methods have been presented that enable the design morphology and control of interesting virtual creatures/robots [129, 130]. In [131] also aspects of the development and production have been considered.

*Machine Learning.* Deep artificial neural network-based approaches have been developed and used for generative design and analysis of materials, biomechanical products [132, 133]. In [89, 90] deep neural networks have been combined with topology in the design and optimization of mechanical structures.

*Reinforcement learning* (RL) approaches have been developed to achieve impressive performance in many applications such as games, control, and simulation-based optimization [60, 134, 135]. Recently also RL methods have been applied in the field of design and engineering, such as drug and circuit design. [136, 137]. A review of advances in reinforcement learning is provided in [135, 138, 139].

*Evolutionary Computing and nature-inspired algorithms* have been used in the design and optimization of software and mechanical systems [90, 91, 119, 140].

*Fuzzy logic* approaches enable the consideration of uncertainties in decision-making and have been used in safety engineering and inference systems [51, 141].

## 4 Discussion and Perspectives

### 4.1 Mind the gap: intelligent systems for design, engineering, and development

Contemporary design, engineering, and development paradigms are still rather human-centered in the execution stage. In a nutshell: engineering development processes are generally executed by a collective network of human agents that drive and control a wide variety of computational tools and automated workflows. In conventional tool-based engineering development paradigms, the involved "narrow" AI agents are rather passive, and require well-defined problems as well as pre- and postprocessing by human agents. Many of the essential activities in design, engineering, and development processes involve aspects of intelligence (e.g., flexibility, adaptivity, problem decomposition, learning, planning, and resource allocation) that are currently still performed and provided to the process by the human agents in the loop using: intuition, experience, reasoning, heuristics, and creativity. Improved understanding and automation of these and similar qualities and capabilities require further interdisciplinary research and progress.

**Towards Computational Rational Processes: interdisciplinary paradigms**
The framework of computational rationality [42, 61] aims to unify the fields of AI, cognitive science, and neuroscience with the goal to exploit synergies in improving the understanding of decision-making and problem solving considering conditions with limited resources for reasoning. In the context of design, engineering, and development processes, problem-solving and decision-making not only involves CR but also intersects with fields such as Design, Engineering Science, Operations Research, Systems Engineering, AI, Computational Physics, Uncertainty Quantification,

Optimization, and Control. A joint framework of Computational Rational design (CRd) Engineering (CRE) and Development (CRD) could bring insightful and rewarding synergies in research and development among all of the involved fields. Besides the economic and technological incentives, there is an abundance of possibilities to collect data and feedback from trained and experienced human agents in the respective fields. The central research goal of CRX is to understand and improve how the decisions, policies, agents, organizational structures with the highest expected utility of the overall process X, given the available resources can be identified and realized. The objectives can go beyond increased understanding and automation of individual human-level capabilities and include aspects related to collective human intelligence and AI-human hybrid intelligence. Human intelligence has been described using agent-based models as a "Society of Mind" in [125]. Improved understanding of complex (engineering) processes involving collective intelligence over cooperative agents requires an inter- or even a transdisciplinary approach and a common vocabulary [142].

*Technical goals and perspectives: Computational Intelligent Driven Development.*
Computationally intelligent systems with higher-level competencies could increase the overall capability and efficiency of design, engineering, and development processes. Besides the current trends in the development of a diversity of AI-agents for specific narrow tasks, it could be rewarding to set goals towards the realization of composite intelligent systems that have the capabilities to perform higher-level tasks and which could eventually drive complex design, engineering and development processes.

Computational Intelligence-Driven Engineering (CIDE) and Development (CIDD) could serve as technical goals towards the automation of engineering and development beyond the current state-of-the-art tool-based "computer-aided" approaches. With CIDE as an initial mid-to-long-term milestone with a focus on automated and autonomous engineering design. CIDD could be a next long-term milestone, additionally including further consideration of a wider range of realization aspects such as the engineering of the manufacturing process and extended product life-cycle impact factors. The development of intelligent systems that are able to "drive" engineering and development processes requires more than just connecting the many narrow-capability agents together in a workflow. Although much can be learned from automated manufacturing systems developed using the industry 4.0 paradigm, the processes and tasks in design engineering and development are more complicated and complex and require the collective of agents to work as an integrated hierarchical system to handle demanding interactive higher-level cognitive tasks. Agents or systems that are more flexible with increasing capabilities in areas such as: problem recognition, problem decomposition or disentanglement, adaptivity, planning and resource allocation, method selection, cooperation, self-reflectivity, and learning are therefore needed.

*Scientific goals and perspectives: Computational Rational Development*
Computational Systems and agents that are can address higher-level and complex engineering development tasks are still an open challenge in science, research, and technology. History indicates that systems are generally realized with increasing levels of complexity when considered functionally and chronologically. Therefore, targets and progress in the direction of systems and agents for gradually increasing levels of complexity and generality are not only of technological importance but could also contribute towards Artificial General Intelligence (AGI). The scientific challenge of CRD goes beyond the technical goal of establishing programmed or trained learning systems that can deal with specific types of complex engineering development tasks, but the overall aim is to establish the frameworks, theories, and methods that enable the realization of intelligent systems that are capable of higher-level tasks of increasing complexity that feature aspects of development. Understanding and realizing intelligent systems that are capable of causal inference (concluding how things are and how they will be) [143] is an important step towards systems that can grasp features of development (realizing how desirable things that have never been could be achieved). Besides the challenges of realizing such systems also aspects of safety and ethics require research consideration [144, 145]. Both inductive research with reasoning and generalization from the specific, as well as deductive research with reasoning from general theories, can be valuable to understand and create the next generation of intelligent systems. It could therefore be beneficial to establish transdisciplinary research frameworks and programs with the goal to increase the understanding of computationally rational decision-making and problem-solving for complex engineering development tasks and processes by intelligent systems with bounded resources.

### 4.2 Open challenges and prospective research directions

Improved understanding and the realization of intelligent systems for design, engineering, and development involve a variety of open multidisciplinary challenges at different process levels:

1. **Domain knowledge, problem specification, and description:** Improved methods to formalize and describe the various decision-tasks, activities, environments, and resources that typically occur in engineering development are necessary.
2. **Task and problem decomposition and recognition:** Research on methods for the characterization, decomposition, categorization, and recognition of tasks and decision problems in sub-tasks/problems.
3. **Policy modeling and evaluation:** Development of methods for the estimation and description of the expected performance, resource requirements, and costs for the different available solution procedures and strategies for the overall and sub-problems, under consideration of the available resources and the involved uncertainties.
4. **Policy selection, planning, and resource allocation:** The endowment of agents or agent-based systems with capabilities for meta-level reasoning regarding policy selection, planning, and resource allocation based on systematic evaluation of the sub-problems.
5. **Adaptive reflective agents:** Improvement of methods to enable agents to reflect their true performance after execution w.r.t. their estimated performance in order to update and learn and performance estimates for policy selection.

6. **Organizing the society of mind:** Development of improved methods to link, combine and organize "narrow" AI Systems together, in ways such that the efficiency or capabilities of the integrated system exceed those of the separate systems.
7. **Information representation and communication:** Investigation and development of effective representations and/or languages to store and communicate: problems, solution procedures, and results in ways that enable recognition, generalization, and adaptation for future tasks and problems.
8. **Language, interaction, and communication**: Development of ways to improve human-machine and machine-human interactions. Not only taking into account communication interfaces but also the information, structure, language and context which is being communicated.
9. **Education:** Cross-disciplinary education and training in AI, design, engineering, and related fields to empower the capabilities of human agents to develop and improve automation systems.

## 5   Concluding Remarks

In order to make progress towards intelligent systems which are able to efficiently realize high-level design, engineering, and development processes, it is necessary to increase the understanding of computational rationality in the context of the complex hierarchically structured task and decision environments occurring in these application domains. To effectively increase the required understanding of the many involved factors, the knowledge and research from various disciplines could be exploited and explored in the scope of transdisciplinary research frameworks such as CRD. This paper highlighted important contributions from various research disciplines, focusing on their intersections related to problem-solving and decision-making processes in design, engineering, and development. Based on the presented mini-review, specific open challenges have been identified, and a road map of future research directions through an interdisciplinary research framework is presented. The overall objective of this contribution was, however, not to restrict future research to specific directions but to motivate and stir up an interdisciplinary discussion and movement to set challenging targets and initiate innovative research. The presented perspectives could extend Herbert Simon's "science of design" [79], towards a science of systems that purposefully design, engineer, and develop.